\documentclass[prb,aps,showpacs,twocolumn,amsmath,amssymb,lengthcheck]{revtex4}
\usepackage{epsfig}
\usepackage{bm}

\begin{document}

\title{Extraordinary wetting phase diagram for mixtures of Bose-Einstein condensates}

\author{{\small J.O. Indekeu and B. Van Schaeybroeck}}
\address{{\small\it Laboratorium voor Vaste-Stoffysica en
Magnetisme,}\\{\small\it Celestijnenlaan 200 D, Katholieke
Universiteit Leuven}, {\small\it B-3001 Leuven, Belgium.}}

\date{\small\it 23 July, 2004}

\begin{abstract}

The possibility of wetting phase transitions in Bose-Einstein
condensed gases is predicted on the basis of Gross-Pitaevskii
theory. The surface of a binary mixture of Bose-Einstein
condensates can undergo a first-order wetting phase transition
upon varying the interparticle interactions, using, e.g., Feshbach
resonances. Interesting ultralow-temperature effects shape the
wetting phase diagram. The prewetting transition is, contrary to
general expectations, not of first order but critical, and the
prewetting line does not meet the bulk phase coexistence line
tangentially. Experimental verification of these extraordinary
results is called for, especially now that it has become possible,
using optical methods, to realize a planar ``hard wall" boundary
for the condensates.

\end{abstract}


\maketitle

In this Letter we pose and answer theoretically the following
fundamental questions. Is a wetting phase transition possible in
ultra-cold dilute gases which undergo Bose-Einstein condensation
(BEC)? How many species of atoms are needed to compose two
coexisting phases with an interface? What is the nature of the
surface or ``wall" at which these phases are ``adsorbed"? If a
wetting transition occurs, is its character first-order or
critical? What is the nature of possible prewetting phenomena away
from bulk two-phase coexistence? We focus on the essential physics
in the application of wetting theory to BEC and give details of
the calculations elsewhere \cite{VSI}.

The simplest system, a one-component gas, is insufficient for
studying wetting transitions, because the condensate fraction and
the normal fraction are fully mixed in position space. Spatial
segregation is only possible in an external potential (e.g.,
gravity) \cite{Huang}. Without this potential an interface between
normal fraction and condensate does not exist, and therefore the
essential interfacial tension is missing.

The next simplest BEC system is a two-component gas. Binary
mixtures of trapped BE condensates of alkali-metal atoms have
received much attention, experimentally \cite{Hall,Stenger} and
theoretically \cite{Ho,Ao,Svid}, since the seminal predictions
concerning their phase behavior by Ho and Shenoy \cite{Ho}.
Different degrees of spatial segregation of two pure-component
condensates at two-phase coexistence, and the possibility of a
symmetric-asymmetric configurational transition in a trap, have
been elucidated by Ao and Chui \cite{Ao} and Svidzinsky and Chui
\cite{Svid}.

We exploit these findings in the context of wetting phenomena
\cite{Cahn,NF,Ross,Bonn}, and point out the existence of a surface
phase transition from partial to complete wetting, upon varying
the intra- or inter-species interactions in mixtures of BE
condensates. The surface consists of a planar ``hard wall", at
which the condensate wave functions vanish.

The basic physics of wetting is best indicated by invoking the
familiar energy balance known as Young's law,
\begin{equation}
\gamma_{_{W1}} = \gamma_{_{W2}} + \gamma_{_{12}}\cos\theta
\end{equation}
where $\gamma_{_{Wi}}$ is the surface (free) energy of a phase of
pure component $i$, $\gamma_{_{12}}$ is the interfacial tension
between pure phases $1$ and $2$, and $\theta$ is the contact angle
with which the 1-2 interface meets the surface. One of the
components, say $2$, is preferentially adsorbed at the surface, so
that $\gamma_{_{W1}} > \gamma_{_{W2}}$. The surface phase
transition from {\em partial wetting} ($\theta >0$; Figure 1a) to
{\em complete wetting} ($\theta = 0$; Figure 1b) is then the
dramatic phenomenon in which, at the surface, pure phase 1 is
displaced by a macroscopically thick wetting layer of pure phase
2. Thus, the $W1$ surface is replaced by a $W2$ surface in
parallel with a 1-2 interface. The physical implications of the
singularity structure around wetting transitions have been the
subject of impressive experimental \cite{Bonn},
theoretical\cite{Boulter} and simulational \cite{Binder} research.

\begin{figure}[h]
   \epsfig{figure=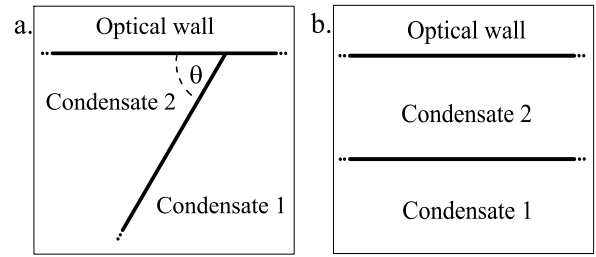,angle=0,width=220pt}
       \caption{a) Partial wetting. The interface between the two condensates makes
       a finite contact angle with the wall. b) Complete wetting. A macroscopic layer of condensate 2
       intrudes between the wall and condensate $1$.
         }
\end{figure}

In view of the ultra-low (typically nanoKelvin) temperature needed
for BEC in dilute gases, the quantum mechanical many-body theory
at $T=0$ is appropriate. For weakly interacting gases the
Bogoliubov mean-field theory of BEC \cite{Huang,Fetter,Pita} is
justified. For our inhomogeneous binary mixture one obtains two
coupled Gross-Pitaevskii (GP) equations for the condensate order
parameters $\psi_1= F_1 e^{i\chi_1}$ and $\psi_2= F_2
e^{i\chi_2}$. We recall that $\psi$ is the ground state
expectation value of the Boson field operator. It can be chosen
real here, since stationary (non-flowing) condensates are assumed,
so $\chi_1 = \chi_2 = 0$.

We adopt the standard geometry for wetting and confine the atoms
to the half-space $z>0$. The relevant surface is the $z=0$ plane.
It is then natural to employ the grand canonical ensemble, with
chemical potentials $\mu_1$ and $\mu_2$ \cite{note1}. The GP
equations are the Euler-Lagrange equations of the grand potential
functional

\begin{align}
\Omega [\psi_1,\psi_2]  = \int _{z\geq 0}\text{d}{\bf r}\,&
\sum_{i=1,2} \left ( \psi_i({\bf r})^* \left [ -
\frac{\hbar^2}{2m_i} {\bf \nabla}^2 - \mu_i \right]\psi_i({\bf r})
 \right.\nonumber\\ &\left.+ \frac{G_{ii}}{2} | \psi_i({\bf r})|^4
\right ) + G_{12} |\psi_1({\bf r})|^2 |\psi_2({\bf r})|^2
\end{align}

The sum is over the two species, with atomic masses $m_i$. The
repulsive interactions $G_{ij}>0$ are related to the $s$-wave
scattering lengths $a_{ij}$ through $G_{ii} = 4\pi \hbar^2
a_{ii}/m_i$ for like particles and $G_{12} = 4\pi \hbar^2
a_{12}(1/m_1+1/m_2)$ for unlike particles. For alkali-metal atoms,
$a \approx 10^2 $\AA.

The confining potential is taken to be a hard wall at $z=0$, so
that
\begin{equation}
\psi_1(x,y,0)=\psi_2(x,y,0)=0
\end{equation}
The closest experimentalists have come to make a hard wall is to
use an evanescent wave, blue-detuned, extending from within a
planar prism \cite{Rychtarik}. The turn-on distance of this
optical wall is typically only about $\lambda/2\pi \approx 80 nm$.
To contain the atoms, a conventional quadratic-confining magnetic
trap can be added, so that a ``quasi-square" exponential potential
results for $z<0$, and the usual $ax^2+by^2+cz^2$ for $z>0$. In
order for volume forces due to a non-uniform external potential to
be negligible compared to surface forces governing wetting, the
harmonic potential must be sufficiently flat-bottomed near ${\bf
r}=0$. For a characteristic length of the order of microns or
longer we can ignore the harmonic potential in the calculations
and assume translational invariance of $\psi$ in $x$ and $y$
directions.

The known condition for bulk phase separation of the binary
mixture is that the unlike particles repel each other more
strongly than the like ones on average, $K \equiv G_{12}/\sqrt{
G_{11}G_{22}} > 1$. Otherwise, a single mixed phase results. Bulk
phase coexistence requires equal pressure, $P_1 = P_2$, in the two
phases, with $P_i = \mu_i/2G_{ii}$. A {\em bulk triple point} with
coexisting mixed and pure phases 1 and 2 is found for $K=1$. This
triple point bears a reminiscence to a critical point in that the
interfacial tension between pure phases 1 and 2 vanishes. However,
the phases themselves remain distinct.

The bulk condensate number densities in the pure phases are
$\rho_{0i} \equiv |\psi_i(\infty)|^2 = \mu_i /G_{ii} $. A
``healing length" $\xi_{i} = \hbar/\sqrt{2 m_i \rho_{0i} G_{ii}}$
characterizes the recovery distance of the order parameter from a
disturbance, e.g., a surface where $\rho = 0$. A typical value is
$\xi \approx 10^3$\AA. The two surface grand potentials
$\gamma_{_{Wi}}$ are proportional to the product of the pressure
and this healing length \cite{Fetter}, $\gamma_{_{Wi}}= (4
\sqrt{2}/3) P \xi_i$, which defines a surface ``thickness". For
strongly repulsive interactions and/or high densities, the healing
length of a condensate is short, and its surface tension is low,
so that it prefers to be near the surface. This suggests to define
a ``surface field" proportional to the difference $\xi_1 - \xi_2$.

The second important length is the ``penetration depth" $\Lambda$,
characterizing the distance over which one condensate decays or
``penetrates"  into the other at the 1-2 interface \cite{Ao},
$\Lambda_i = \xi_i/\sqrt{K-1}$. A typical value is $\Lambda
\approx 1\mu m$. This length diverges when approaching the triple
point. Ao and Chui discuss the following two limits for which we
determine the wetting behavior.

{\bf A) Strong segregation limit: Partial Wetting}\\
In the limit $\Lambda \ll \xi$ or $K \rightarrow \infty$ the
condensates show almost no mutual penetration. The interface then
consists of two surfaces, one where condensate 1 decays to zero
over the healing length $\xi_1$, and an adjacent one where
condensate 2 vanishes over $\xi_2$. The interface thickness is
thus $\xi_1 + \xi_2$ and its tension is approximately
$\gamma_{_{12}} \approx \gamma_{_{W1}} +\gamma_{_{W2}}$. There can
be no complete wetting, since the energy cost of an interface is
too high to be able to satisfy Young's law with $\theta = 0$.

{\bf B) Weak segregation limit: Complete Wetting}\\
In the experimentally more relevant limit $\Lambda \gg \xi$ or $K
\rightarrow 1$ the mutual penetration of the condensates leads to
mixing on the scale of $\Lambda_1 + \Lambda_2$. In spite of its
huge thickness, the interface has a low tension, since the energy
per unit volume scales as\cite{Ao,VSI} $(\xi/\Lambda)^2P$. As a
result\cite{Barankov}, $\gamma_{_{12}} \propto P (\xi_1 + \xi_2)
\sqrt{K-1}$. The interfacial tension vanishes at the triple point
with a square-root singularity. For Young's contact angle we
obtain
\begin{equation}\label{phasebound}
\cos\theta = const. \left ( \frac{\xi_1 - \xi_2}{\xi_1 +
\xi_2}\right ) (K-1)^{-1/2}
\end{equation}
This implies that a wetting transition is unavoidable upon
approach of the triple point. Indeed, the wetting phase boundary
implied by (\ref{phasebound}) is given by
\begin{equation}
(\xi_1/\xi_2 - 1) \propto (K-1)^{1/2}
\end{equation}

\begin{figure}[h]
   \epsfig{figure=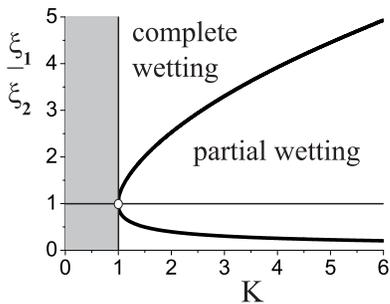,angle=0,width=145pt}
       \caption{Wetting phase diagram in the plane of surface field $(\xi_1-\xi_2)/\xi_2$ and relative
       interaction strength $K$. At bulk two-phase coexistence ($K>1$) a first-order
       phase boundary separates partial wetting from complete wetting, for $\xi_1/\xi_2>1$.
       For $\xi_1/\xi_2 <1$ the roles of the condendates are interchanged and one may use
       ``drying" in place of ``wetting". The phase boundary is
       parabolic near the triple point $(1,1)$.
        }
\end{figure}

A full numerical computation \cite{VSI} corroborates the physical
insights gained so far, and leads to the wetting phase diagram at
bulk two-phase coexistence shown in Fig.2. The wetting transition
is of {\em first order}. On the phase boundary the grand potential
of partial wetting crosses that of complete wetting, and both
states have metastable continuations. Interestingly, at the
wetting transition the grand potential is degenerate: all wetting
layers have the same energy, regardless of the layer thickness.
This is akin to what happens at the wetting transition in an Ising
model at $T=0$, where the interface consists of a plane of broken
bonds at an arbitrary distance from the surface. The usual
entropic repulsion between the surface and the interface due to
interfacial capillary wave fluctuations is absent at $T=0$.

Yet, in contrast with Ising spins on a lattice, the continuum
field theory we are dealing with allows one to ``inflate" a
wetting layer by varying its shape and thickness continuously
between infinitesimal and macroscopic, at {\em constant} grand
potential. A remarkable consequence of this is the coincidence, at
bulk two-phase coexistence, of {\em nucleation} and wetting
transitions \cite{VSI}. For illustration, Figure 3 shows two order
parameter profiles for parameters $K$ and $\xi_{1}/\xi_2$ on the
wetting line.
\begin{figure}[h]
   \epsfig{figure=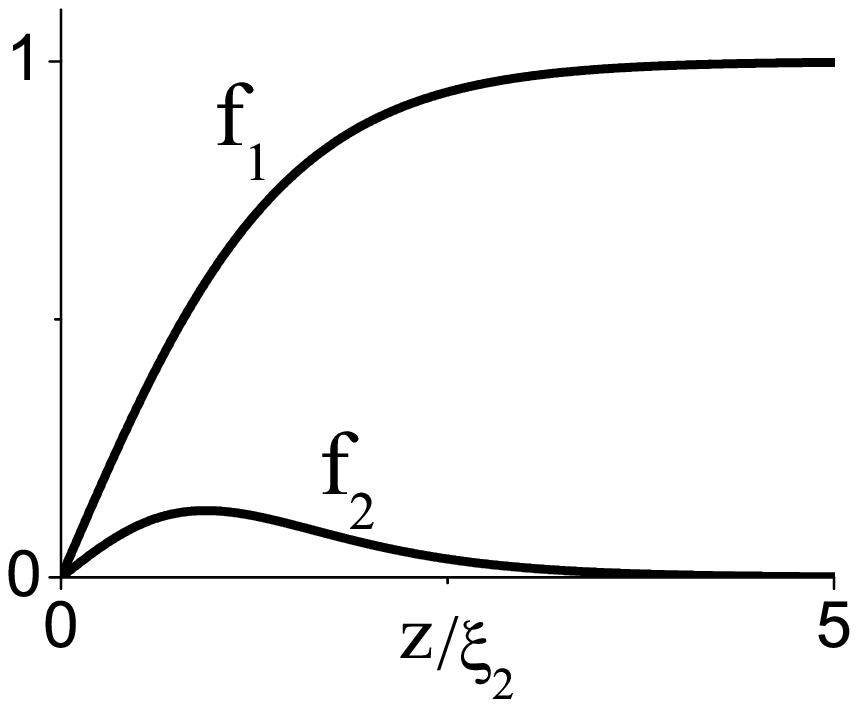,angle=0,width=120pt}
   \epsfig{figure=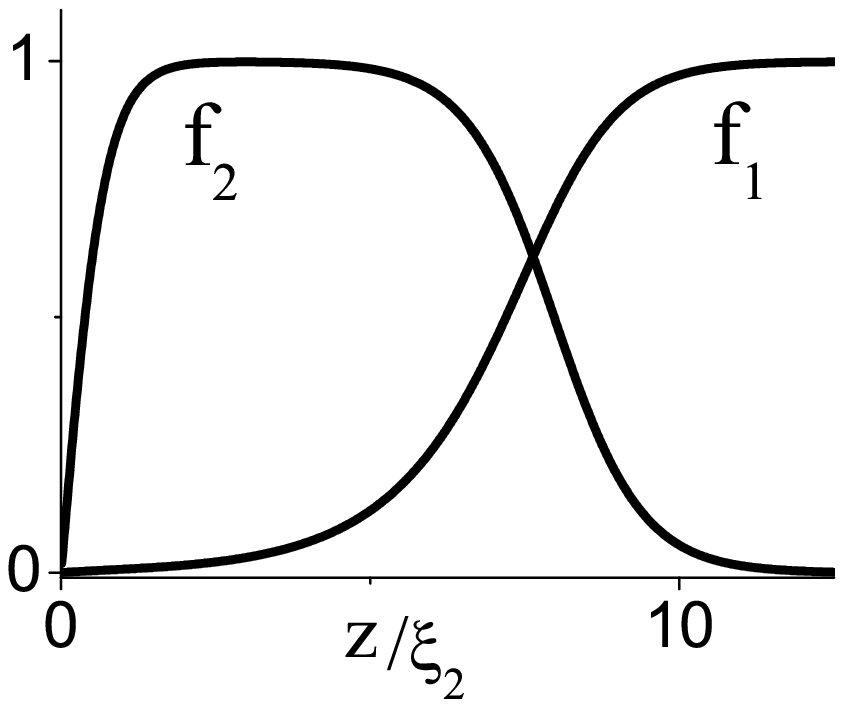,angle=0,width=120pt}
       \caption{Inflation of a wetting film of condensate 2 at the surface of condensate 1.
       Normalized order parameters,
       $f_{i} \equiv F_{i}/\sqrt{\rho_{0i}}$, are shown for two
       surface states with the same grand potential, at first-order wetting ($K=1.5$,
$\xi_{1}/\xi_{2}=2$).
       }
\end{figure}

A crucial physical question is how the surface field and/or the
relative interaction strength can be tuned experimentally.
Importantly, these variables can be expressed in terms of the
atomic masses and scattering lengths alone, and are therefore
fully microscopic. Fesh\-bach resonances allow manipulation of the
scattering length(s) over more than an order of magnitude
\cite{Inouye} and are a suitable tool for exploring the phase
diagram.

Away from bulk two-phase coexistence, a first-order wetting
transition normally possesses an extension into the bulk one-phase
region where the wetting phase is metastable. If this extension
exists, general arguments show that it must also be of
first-order, at least from the wetting point up till a prewetting
(multi-)critical point in the phase diagram. Furthermore, the
prewetting line must meet the bulk two-phase coexistence line
tangentially \cite{Hauge}. To our knowledge, in all systems that
have been observed or predicted to undergo a first-order wetting
transition, ranging from classical adsorbed fluids \cite{Bonn} or
superfluids \cite{Ross}, over magnetic systems \cite{Binder}, to
type-I superconductors \cite{IvL}, the prewetting transition is
also of first order. Yet, mixtures of BE condensates provide us
with an exception, as we now show.

Figure 4 shows the prewetting phase diagram in the space of ``bulk
field" and relative interaction strength $K$. The bulk field which
can drive the mixture away from bulk two-phase coexistence is
proportional to the pressure difference $P_1 - P_2$. The
combination $\sqrt{P_1/P_2}-1$ serves the same purpose. Note that
pure phase 1 (2) is stable in bulk for $P_1 > (<) P_2$ and that
the mixed phase is stable for $ K < $ min$ \{
\sqrt{P_1/P_2},\sqrt{P_2/P_1}\}$. The bulk triple point is at
$(1,1)$. For a given value of the surface field, i.e., for fixed
$\xi_1/\xi_2$, the wetting point $W$ and the prewetting line are
shown. When, along the path indicated by arrows, the prewetting
transition is crossed, an infinitesimal film of condensate 2 is
nucleated at the surface of condensate 1. Moving further to the
left, the thickness of this film grows and, upon approach of
two-phase coexistence (at $P_1=P_2$), it diverges logarithmically.
This slow divergence is expected for the approach to {\em complete
wetting} in systems with short-range interactions. The
extraordinary physics emanating from this phase diagram concerns
i) the second-order or ``critical" nature of the entire prewetting
line, and ii) the fact that the prewetting line meets bulk
two-phase coexistence at a finite angle (clearly seen from the
mathematical continuation along the dashed line).

\begin{figure}[h]
   \epsfig{figure=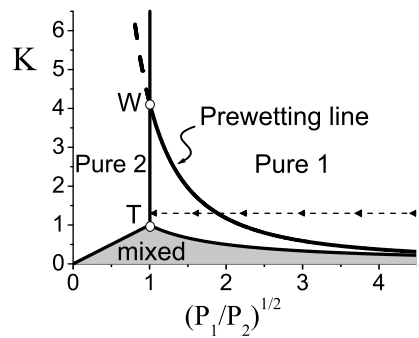,angle=0,width=155pt}
       \caption
       {
       The phase diagram for prewetting at $\xi_1/\xi_2=4$, in the space of relative
       interaction strength $K$ and bulk field $\sqrt{P_1/P_2}-1$.
       The bulk phases ``pure 1", ``pure 2" and ``mixed" coexist
       at the triple point $T$. Bulk two-phase coexistence of the
       pure condensates is along the line $P_1=P_2$, on which the
       first-order wetting transition $W$ is indicated. The second-order prewetting phase
       boundary intersects
       the bulk transition at $W$. The line with arrows indicates
       a possible approach to complete wetting.
             }
\end{figure}

If there is nothing wrong with the general thermodynamic arguments
concerning prewetting, how is this anomaly possible? There are two
premises in the general reasoning, which are {\em not} fulfilled
here. Firstly, at this first-order wetting transition there is no
abrupt jump of the film thickness between a microscopic and a
macroscopic value, but all intermediate values are equally stable
due to the grand potential degeneracy. Since there is no interface
potential barrier, for consistency the prewetting transition must
be continuous instead of first-order, and we find that it is.
Secondly, since this prewetting transition is a second-order
nucleation transition, unlike for ordinary prewetting, the
difference in adsorption on either side of the transition does not
diverge upon approach of the wetting transition at $W$, and the
condition for a tangential meeting is not met. Consistently, we
find an intersection at a finite angle.

Within the wetting phenomenology we reinterpret the interesting
symmetric-asymmetric (SA) transitions predicted for an axially
symmetric square-well trap \cite{Ao}, and for a conventional
quadratic magnetic trap \cite{Svid}. In the symmetric (S) state
the condensates form concentric clouds. This corresponds to
complete wetting (Fig.1b), with, e.g., a $W2$ surface and an
embedded 1-2 interface in a ``cherry" configuration. This state
may exchange stability with one (A) which breaks the trap
symmetry, with a planar 1-2 interface cutting across the system.
This is partial wetting (Fig.1a).

For a square-well trap, a surface energy balance determines the SA
transition \cite{Ao}. However, it is not governed by Young's law,
but by finite-size effects, cylindrical geometry and particle
number conservation. Consequently, the contact angle at the SA
transition does not tend to zero continuously as in the {\em bona
fide} wetting transition. Instead, it jumps from a finite value
$\theta_{SA}$ to zero. For a quadratic trap, surface and volume
contributions are entangled in the energy balance \cite{Svid}. In
the S state the component with the largest self-repulsion is on
the outside. For sufficiently small trap radius, an A state is
favored, to relieve the high capillary pressure across the curved
1-2 interface in the S state. The wetting interpretation is
complicated here due to the spatially varying external potential
and the absence of an articulated surface.

In conclusion, in view of the extraordinary ultralow-temperature
effects in the predicted wetting phase diagram for binary mixtures
of Bose-Einstein condensed gases, we advocate an experimental
study of wetting and prewetting in a trap which is suitable for
observing true wetting singularities. The confinement should
consist of a half-space with one planar ``hard wall", i.e., a
steep repulsion with a turn-on length smaller than the healing
lengths and penetration depths of the condensates. A mild
conventional trap may be used to keep the gas near the wall. The
wetting transitions can be induced by tuning the interatomic
scattering lengths using Feshbach resonances.

We thank Henk Stoof, Lev Pitaevskii and Volker Weiss for helpful
discussions at the start of this work. We are grateful to Eric
Cornell and Jacques Tempere for helpful remarks regarding
experimental ``hard walls". This research is supported by grant
FWO-G.0222.02.

\newpage


\begin{thebibliography}{99}
\bibitem{note1}
Calculations for the canonical ensemble, with constrained numbers
of condensate atoms $N_1$ and $N_2$ were also performed
\cite{VSI}. Canonical surface and interfacial energies are smaller
than their grand-canonical counterparts by a factor of 4. The
contact angle and wetting phase boundary are ensemble-independent.
\end{thebibliography}
\end{document}